# $Mn_3O_4$-Graphene Hybrid as a High Capacity Anode Material for Lithium Ion Batteries

Hailiang Wang,[†,§] Li-Feng Cui,[‡,§] Yuan Yang,[‡] Hernan Sanchez Casalongue,[†] Joshua Tucker Robinson,[†] Yongye Liang,[†] Yi Cui[*,‡] and Hongjie Dai[*,†].

[†]*Department of Chemistry and Laboratory for Advanced Materials, and* [‡]*Department of Materials Science and Engineering, Stanford University, Stanford, CA 94305, USA.*

[§] These two authors contributed equally to this work

hdai@stanford.edu and yicui@stanford.edu

**Abstract** We developed two-step solution-phase reactions to form hybrid materials of $Mn_3O_4$ nanoparticles on reduced graphene oxide (RGO) sheets for lithium ion battery applications. $Mn_3O_4$ nanoparticles grown selectively on RGO sheets over free particle growth in solution allowed for the electrically insulating $Mn_3O_4$ nanoparticles wired up to a current collector through the underlying conducting graphene network. The $Mn_3O_4$ nanoparticles formed on RGO show a high specific capacity up to ~900mAh/g near its theoretical capacity with good rate capability and cycling stability, owing to the intimate interactions between the graphene substrates and the $Mn_3O_4$ nanoparticles grown atop. The $Mn_3O_4$/RGO hybrid could be a promising candidate material for high-capacity, low-cost, and environmentally friendly anode for lithium ion batteries. Our growth-on-graphene approach should offer a new technique for design and synthesis of battery electrodes based on highly insulating materials.

Nanomaterials of metal oxides, such as $Co_3O_4$,[1,2] $SnO_2$,[3] $FeO_x$[1,4] and NiO[1] have been intensively studied as anode materials for lithium ion batteries (LIBs) aimed at achieving higher specific capacities than graphite. For instance, nanocrystals of $Co_3O_4$ have been synthesized for LIB anodes with specific capacities ~2X of that of graphite, affording LIBs with higher energy density.[1,2] Compared to $Co_3O_4$, $Mn_3O_4$ is an attractive anode material for LIBs due to the high abundance of Mn, low cost and environmental benignity. However, little has been done thus far in utilizing $Mn_3O_4$ as an anode material[5,6] partly due to its extremely low electrical conductivity (~$10^{-7}$-$10^{-8}$ S/cm) limiting its capacity ≤ ~400mAh/g even with Co-doping,[6] well below the theoretical capacity of ~936mAh/g. It is highly desirable to increase the capacity of $Mn_3O_4$ by electrically wiring up $Mn_3O_4$ nanoparticles to an underlying conducting substrate.

Graphene is an excellent substrate to host active nanomaterials for energy applications due to its high conductivity, large surface area, flexibility and chemical stability. We have shown recently that the specific capacitance and power rate of $Ni(OH)_2$ could be enhanced by growing $Ni(OH)_2$ nanoplates on graphene sheets, useful for supercapacitor applications.[7] $Co_3O_4$-graphene and $Fe_3O_4$-graphene composite materials have also been recently prepared as anode materials of LIBs,[8] with comparable performance to previously reported $Co_3O_4$ and $Fe_3O_4$ based anodes.[2a,2c,4a,4c] It is more challenging to explore $Mn_3O_4$-graphene hybrid materials due to the much lower conductivity of $Mn_3O_4$ than $Co_3O_4$ and $Fe_3O_4$. Here, we report a two-step solution phase method of growing $Mn_3O_4$ nanoparticles on graphene oxide (GO) to form a $Mn_3O_4$-reduced GO (RGO) hybrid material. The $Mn_3O_4$/RGO hybrid afforded an unprecedented high capacity of ~900mAh/g based on the mass of $Mn_3O_4$ (~810mAh/g based on the total mass of the hybrid) with good rate capability and cycling stability. This growth-on-graphene approach would be very useful in boosting the electrochemical performance of highly insulating electrode materials.

$Mn_3O_4$ nanoparticles were grown on GO sheets by a two-step solution phase reaction scheme [Figure 1, also see Supporting Information (SI) for details] recently developed for the synthesis of $Ni(OH)_2$/graphene hybrid.[7] GO used in this work was prepared by a modified Hummers method (SI),[9] in which a 6 times lower concentration of $KMnO_4$ was used (see SI) to result in GO sheets with lower oxygen content than Hummers' GO (~15% vs. ~30% measured by XPS and Auger spectroscopy).

In the first step, hydrolysis of $Mn(CH_3COO)_2$ in a GO suspension with a 10:1 *N, N*- dimethylformamide (DMF)/$H_2O$ mixed solvent at 80°C afforded dense and uniform coating with small and poorly crystalline precursor nanoparticles on GO (see SI and Figure S1). The mixed solvent and low temperature were important to control the hydrolysis reaction at a low rate to avoid particle formation in free solution. It is possible that nanoparticles grown this way were anchored covalently to GO through functional groups such as carboxyl, hydroxyl and epoxy groups on the surface of GO sheets.

The product from the first step was then transferred to de-ionized water and treated in hydrothermal conditions at 180°C for 10 hours, which afforded well crystallized $Mn_3O_4$ nanoparticles uniformly distributed on RGO sheets (Figure 1), with ~10*wt%* RGO in the hybrid material. Through the synthesis, the starting GO sheets evolved into reduced graphene oxide due to the hydrothermal treatment at 180°C, as evidenced by the red shift of the absorption peak in UV-Vis spectra (Figure S2).[10] The size of the $Mn_3O_4$ nanoparticles was ~10-20nm as revealed by the scanning electron microscopy (SEM) and transmission electron microscopy (TEM) images (Figure 1b, 1d), which was consistent with the width of the diffraction peaks in XRD spectrum (Figure 1c). High resolution TEM image (Figure 1e) showed the crystal lattice fringes throughout the entire $Mn_3O_4$ nanoparticle formed on RGO.

The $Mn_3O_4$/RGO hybrid material was mixed with carbon black and polyvinyldifluoride (PVDF) in a weight ratio of 80:10:10 for preparing a working electrode. The electrochemical measurements were carried out in coin cells with a Li foil as the counter electrode and 1M $LiPF_6$ in 1:1 ethylene carbonate (EC) and diethyl carbonate (DEC) as the electrolyte. Figure 2a showed the charge and discharge curve of $Mn_3O_4$/RGO anode for the first cycle at a current density of 40mA/g and cycled between 3.0V-0.1V vs $Li^+$/Li. The capacity corresponding to the voltage range of ~1.2-0.4V of the first $Li^+$ charge curve was due to irreversible reactions between $Li^+$ and RGO and decomposition of the electrolyte solvent forming a solid electrolyte interphase (SEI).[11] The voltage plateau at ~0.4V reflected the Li$^+$ charge reaction: $Mn_3O_4 + 8Li^+ + 8e^- \rightarrow 3Mn(0) + 8Li_2O$.[8b] The discharge curve showed a sloping plateau at ~1.2 V due to the reverse reaction. After several conditioning cycles, the Coulombic efficiency of the coin cell increased to higher than 98% (Figure 2b, 2c), indicating good reversibility of the above conversion reaction between $Mn_3O_4$ and Mn(0). The charge capacity in the range of 1.2-0.4V was mainly due to reduction from Mn(III) to Mn(II), and the 0.4-0.1V range reflected the reduction from Mn(II) to Mn(0).[6] Little charge capacity would be obtained in the rage of 1.2-0.5V if the electrode was cycled between Mn(II) and Mn(0).[6] This further verified the conversion reaction we suggested above.

Figure 2b showed representative charge and discharge voltage profiles of the $Mn_3O_4$/RGO anode at various current densities (see Figure S4 for other data sets). With the increase of current density, the charge potential of $Mn_3O_4$/RGO anode decreased and the discharge potential increased rendering higher overpotential. The cell was first cycled at a low current density of 40mA/g for 5 cycles, where a stable specific capacity of ~900mAh/g (Figure 2c) was obtained based on the mass of the $Mn_3O_4$ nanoparticles in the $Mn_3O_4$/RGO hybrid (~810mAh/g based on the total mass of the hybrid), which was very close to the theoretical capacity of ~936mAh/g for $Mn_3O_4$ based on the conversion reaction above. This is also the highest capacity that has been reported for $Mn_3O_4$ anode materials. To our knowledge, previously, the highest capacity with similar cycle stability for $Mn_3O_4$ anode was ~500mAh/g at ~33-50mA/g.[6] Our $Mn_3O_4$ nanoparticles formed on RGO showed good rate performance as well. The capacity was as high as ~780mAh/g after we increased the current density by 10 times. Even at a high current density of 1600mA/g, the specific capacity was ~390mAh/g, still higher than the theoretical capacity of graphite (~372mAh/g). A capacity of ~730mAh/g at 400mA/g was retained after 40 cycles of charge and discharge at various current densities (Figure 2c), indicating good cycling stability (cycling data up to 150 cycles available in Figure S4).

The high capacity and rate capability, and good cycling stability of our $Mn_3O_4$/RGO hybrid material were attributed to the intimate interaction between the graphene substrates and the $Mn_3O_4$ nanoparticles directly grown on them, which made $Mn_3O_4$ electrochemically active since charge carriers could be effectively and rapidly conducted back and forth from the $Mn_3O_4$ nanoparticles to the current collector through the highly conducting three-dimensional graphene network. The graphene-nanoparticle interaction also afforded a good dispersion of the $Mn_3O_4$ nanoparticles grown on the RGO sheets to avoid aggregation (Figure S5), desired for cycle stability.

In a control experiment, we synthesized free $Mn_3O_4$ nanoparticles by the same two-step method without any graphene added (SI). Although the free $Mn_3O_4$ nanoparticles had similar morphology, size and crystallinity as those formed on RGO sheets (Figure S3), the electrochemical performance of these free $Mn_3O_4$ nanoparticles physically mixed with carbon black was much worse than the $Mn_3O_4$/RGO hybrid material mixed with carbon black. At a low current density of 40mA/g, the free $Mn_3O_4$ nanoparticles showed a capacity lower than 300mAh/g, which further decreased to ~115mAh/g after only 10 cycles (Figure 2d).

In conclusion, by selectively growing $Mn_3O_4$ nanoparticles on graphene oxide to form $Mn_3O_4$/RGO hybrid, we enabled $Mn_3O_4$ approaching its theoretical capacity as an anode material for LIBs. The $Mn_3O_4$-graphene hybrid could be further explored for high-capacity, low-cost and non-toxic anode material for battery applications. Our growth-on-graphene approach should also offer an effective and convenient technique to improve the specific capacities and rate capabilities of highly insulating electrode materials in the battery area.

**Acknowledgement.** This work was supported in part by Office of Naval Research and KAUST Investigator Award.

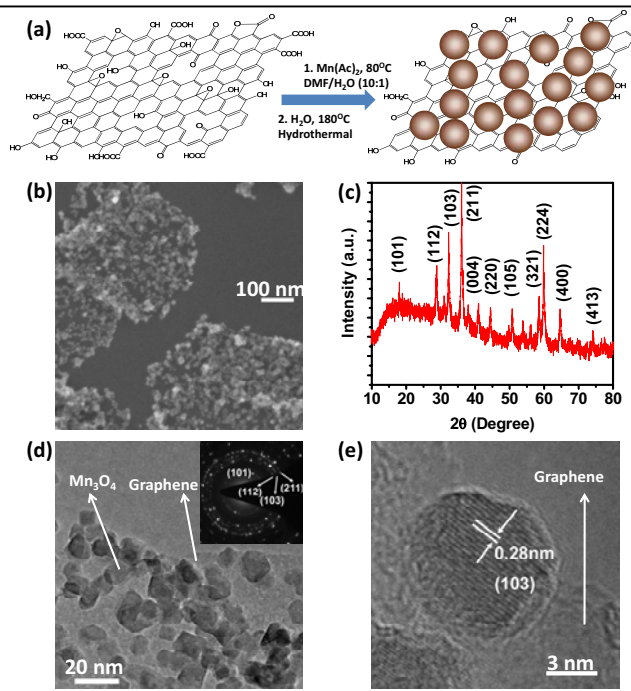

*Figure 1.* Mn$_3$O$_4$ nanoparticles grown on GO. (a) Schematic two-step synthesis of Mn$_3$O$_4$/RGO. (b) SEM image of Mn$_3$O$_4$/RGO hybrid. (c) XRD spectrum of a packed thick film of Mn$_3$O$_4$/RGO. (d) TEM image of Mn$_3$O$_4$/RGO, inset shows the electron diffraction pattern of the Mn$_3$O$_4$ nanoparticles on RGO. (e) High resolution TEM image of an individual Mn$_3$O$_4$ nanoparticle on RGO.

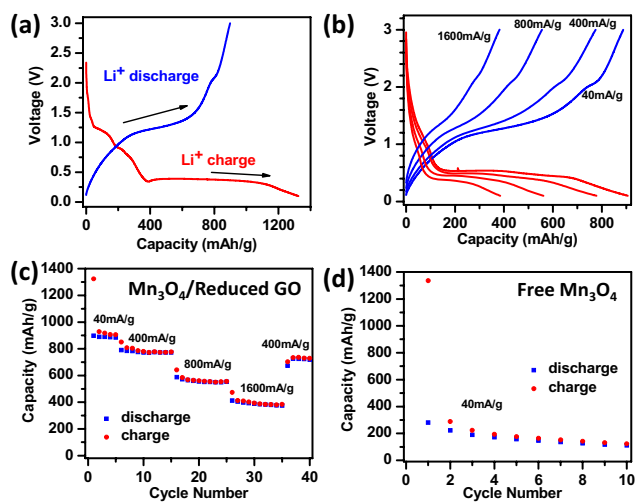

*Figure 2.* Electrochemical characterizations of a half cell composed of Mn$_3$O$_4$/RGO and Li. The specific capacities are based on the mass of Mn$_3$O$_4$ in the Mn$_3$O$_4$/RGO hybrid. (a) Charge (red) and discharge (blue) curves of Mn$_3$O$_4$/RGO for the first cycle at a current density of 40mA/g. (b) Representative charge (red) and discharge (blue) curves of Mn$_3$O$_4$/RGO at various current densities. (c) Capacity retention of Mn$_3$O$_4$/RGO at various current densities. (d) Capacity retention of free Mn$_3$O$_4$ nanoparticles without graphene at a current density of 40mA/g.